\documentclass[prd,aps,superscriptaddress,floatfix,nofootinbib,eqsecnum,notitlepage]{revtex4-1}
\pdfoutput=1

\usepackage{amsfonts}
\usepackage{amsmath}
\usepackage{amssymb}
\usepackage{tikz}
\usepackage{bm}
\usepackage{dcolumn}
\usepackage{graphicx}   
\usepackage[utf8]{inputenc}
\usepackage{latexsym}
\usepackage{rotating}
\usepackage{hyperref}
\usepackage{subfigure}
\usepackage{color}
\usepackage{changes}

\begin{document}

\title{Quantum H\v{o}rava–Lifshitz  Cosmology in the de Broglie-Bohm interpretation}

\author{G.~S.~Vicente}\email{gustavo@fat.uerj.br}
\affiliation{Faculdade de Tecnologia, Universidade do Estado do Rio de
  Janeiro, 27537-000 Resende, RJ, Brazil}

\begin{abstract}

Classical and quantum Friedmann–Lemaître–Robertson–Walker universes filled with non-interacting radiation and dust fluids are considered in the framework of Ho\v{r}ava–Lifshitz gravity theory. 
The Ho\v{r}ava–Lifshitz theory is set in its projectable version and without detailed balance condition.
Canonical quantization is performed in the Wheeler–DeWitt approach of quantum cosmology for a minisuperspace model in the light of de Broglie-Bohm interpretation of quantum mechanics.
The main results are analytical solutions for non-singular quantum bounce and cyclic universes for open and closed spatial sections in terms of the parameters of Ho\v{r}ava–Lifshitz theory.  
\end{abstract}

\maketitle 

\section{Introduction}

The standard model of particle physics successfully describes electro-weak and strong interactions, which operate at the quantum level. 
General Relativity (GR) is the most successful theory of gravitation, which is a classical theory of space-time and matter. 
A first principles description of nature seems to be quantum mechanical, then it is a natural way to assume that gravity must be described by a quantum theory as well.

Applying the canonical quantization scheme to GR, we achieve the so called canonical quantum gravity theory~\cite{DeWitt}, which is a constrained system governed by the Wheeler-DeWitt equation~\cite{QG}. 
The solution of the Wheeler-DeWitt equation yields the quantum state of the 3-geometry and matter fields by a wave functional (on superspace), which describes the whole Universe.  
However, this formulation faces some drawbacks. 
Firstly, we expect that the wave function constrains the dynamics of the Universe, but the absence of a momentum which is canonically conjugate to the time variable implies that the wave function is static. This is called the problem of time~\cite{time}. Secondly, the wave function describes no particular metric, but all spacelike metrics (superspace).
Both these issues compromise the definition and identification of space-time singularities.
Finally, measurement demands a collapse of the wave function by an external observer, which is a feature of the Copenhagen interpretation of quantum theory. This is called the measurement problem.

In order to circumvent these issues, we may consider the de Broglie-Bohm (dBB) interpretation of quantum theory~\cite{Bohm,hollandbook} instead of the orthodox one.
The time degree of freedom can now be defined from matter degrees of freedom if it is described by a hydrodynamical perfect fluid. This can be done using the Schutz formalism~\cite{Schutz}, where the time variable is associated to the fluid’s potentials.
We can also define a metric, which evolves in time according to guidance equations. 
The measurement problem is either eliminated, because in the ontological interpretation the evolution of the Universe is deterministic and do not demand a collapse of the wave function or the action of an external observer.
Therefore, the dBB interpretation of canonical quantum gravity~\cite{Vink:1990fm,Shtanov:1995ie,Goldstein:1999my,PintoNeto:2004uf,Pinto-Neto:2013toa} is a suitable way to establish a quantum cosmology theory. 
In this context, one can define contraction/expansion as well as singularities in space-time.

In order to solve the Wheeler-DeWitt equation in a simple form, we restrict the superspace to the minisuperspace~\cite{ryan,Kuchar:1989tj,Halliwell}, where the degrees of freedom are reduced to a finite number while still preserving the main qualitative aspects of the full picture.
This is a reasonable framework to develop quantum cosmology.

Quantum cosmology models in minisuperspace in the dBB interpretation are vast in the literature~\cite{Vink:1990fm,KowalskiGlikman:1990df,Squires:1992ex,Colistete:1997sf,Kim:1997te,AcaciodeBarros:1997gy,AcaciodeBarros:1998nb,Colistete:2000ix,Pedram:2007mj,Monerat:2010vf,Vakili:2012nh,Das:2013oda,Ali:2014qla,Oliveira-Neto:2014jaa,Peter:2006hx}. 
Particularly, bouncing models are a very relevant feature of quantum cosmology, where quantum effects are responsible for avoiding the big bang (and eventually big crunch) singularity. 
In this context, the authors of Refs.~\cite{Peter:2006hx,AcaciodeBarros:1997gy,Alvarenga:2001nm,PintoNeto:2005gx} present bouncing models for perfect fluids using the Schutz formalism, whereas in Ref.~\cite{Falciano:2007yf} a scalar field is considered as the matter content.
In the case of perfect fluids, a very interesting result is given in Ref.~\cite{PintoNeto:2005gx}, which is the quantum version of the Friedmann-Lemaître-Robertson-Walker (FLRW) universe 
filled with radiation and dust, where radiation dominates around the bounce and dust dominates in the contraction and expansion phases far from the bounce.
Cyclic universe solutions are also present for closed spatial section, where multiple bounces are present.
These results will be very important for this work.

It is important to mention that a cosmological bounce can solve the standard cosmological puzzles~\cite{Peter:2008qz}, like the horizon and flatness problems, and are also robust to generate primordial cosmological perturbations with almost scale invariant spectrum~\cite{Peter:2008qz,Peter:2006hx}.
However, although an inflationary phase can be avoided for these models, in principle it can still happen.
Finally, dark matter has been also considered in this context through the dynamics of a scalar field~\cite{dBBscalarfield}.

Recently, a new theory of gravity was proposed by H\v{o}rava~\cite{Horava:2009uw,Horava:2009if}, which is based on the introduction of an anisotropic scaling between space and time. 
The theory is inspired in the Lifshitz work about critical exponents in phase transitions in the context of condensed matter physics, then referred to as H\v{o}rava-Lifshitz (HL) theory.
The anisotropy is introduced in order to generalize GR in the high energy ultraviolet (UV) scale, while recovering GR in the low energy infrared (IR) scale.
This generalization is due to Lorentz symmetry breaking in the UV scale through a Lifshitz-like process, while this symmetry is preserved in the IR scale.
Due to the anisotropy in space and time, the HL theory is usually  represented in the Arnowitt-Deser-Misner (ADM) formalism~\cite{Arnowitt:1959ah}, which splits the 4-metric $g_{\mu\nu}$ in a 3-metric $h_{ij}$, a shift 3-vector $N_i$ and a lapse function $N$, where the latter parametrizes time.

There are two important issues to consider in HL theory.
Firstly, the variables in the ADM formalism may depend on both space and time. However, H\v{o}rava pointed out~\cite{Horava:2009uw} that the lapse function $N$ should only depend on time, which is reasonable in the cosmological setting. 
This is called the {\it projectable condition}.
On the other hand, some authors have considered the general case in which $N$ depends on both space and time, i.e., a non-projectable condition.
Both projectable and non-projectable theories suffer from problems such as ghost scalar modes and instabilities~\cite{Sotiriou:2010wn,Visser:2011mf,Wang:2017brl}.
However, a consistent projectable HL gravity theory with a local $U(1)$ symmetry can be shown to eliminate the ghost scalar modes~\cite{Horava:2010zj,daSilva:2010bm}, so that instability and strong coupling does not happen in the gravitational sector.
On the other hand, non-projectable HL gravity theories can also be consistent whether the local $U(1)$ symmetry is present~\cite{Zhu:2011yu,Zhu:2011xe,Lin:2013tua} or not~\cite{Blas}, where the latter theory is the so called healthy extension.
A review of the aforementioned issues is given in Ref.~\cite{Wang:2017brl}.
From another point of view, in the cosmological setting, it seems risky to impose that the lapse function only depends on time. However, in this case the Hamiltonian constraint is no longer local but integrated over all spatial volume. This result corresponds to Friedmann equations with an additional cold dark matter-like component~\cite{Mukohyama:2009mz}.
However, in homogeneous models these spatial integrals are simply the spatial volume, then the additional matter content vanishes~\cite{Sotiriou:2009gy,Sotiriou:2009bx}.
These considerations can be extended to the quantum realm in minisuperspace cosmology models~\cite{Bertolami:2011ka,Pitelli:2012sj}, described by the projectable version of the HL theory. Then, it is reasonable to consider the projectable HL in this context.
It is important to mention that the aforementioned theory with local $U(1)$ symmetry can also be applied to the cosmological setting~\cite{Huang:2012ep}.

Secondly, H\v{o}rava also considered a simplification in order to reduce the number of terms of his theory, which is called the {\it detailed balance condition}. 
However, although
the detailed balance is a simplifying assumption, it is not really necessary~\cite{Sotiriou:2009gy,Sotiriou:2009bx}. 
Therefore, in this paper I consider the projectable version of HL theory without detailed balance condition.

Non-singular classical HL cosmological models have been considered in the literature for the matter bounce scenario~\cite{Brandenberger:2009yt}, where the non-singular behavior is due to the presence of spatial curvature.
However, non-singular classical cosmologies can also be achieved by other mechanisms, such as from matter contents~\cite{BounceMatter} and other modified gravity theories~\cite{BounceOthers}. 
On the other hand, additionally to the aforementioned quantum non-singular cosmologies in minisuperspace in the dBB interpretation, the literature is also vast for non-singular quantum cosmological  models~\cite{nonsingularquantum} and also heuristically motivated ones~\cite{Brandenberger:2017pjz}.
Particularly, there is also a great literature for quantum non-singular cosmologies in the framework of Loop Quantum Cosmology~\cite{Ashtekar:2021kfp,ElizagaNavascues:2020uyf,Li:2021mop,Li:2018vzr,Zhu:2017jew}.

H\v{o}rava-Lifshitz quantum cosmology in minisuperspace has already been considered in the literature~\cite{Sotiriou:2009bx,Bertolami:2011ka,Pitelli:2012sj,Vakili:2013wc,Oliveira-Neto:2017yui,Oliveira-Neto:2019tvo,Tavakoli:2021kyc}.
Particularly, exact solutions have been obtained in Refs.~\cite{Vakili:2013wc,Oliveira-Neto:2017yui} for perfect fluids for some values of the equation of state parameter and for some of the parameters of HL theory. 
In this work, I present exact solutions for non-singular universes in the cases of  closed and open FLRW quantum cosmologies in HL theory, where the matter content is composed of non-interacting  radiation and dust fluids.

This paper is organized as follows. 
In Sec.~\ref{FLRWcosmology}, I introduce the FLRW cosmology in HL gravity in the ADM formalism and the Schutz formalism for the matter content.
The gravitational and matter Hamiltonians are presented and the full Hamiltonian for a universe filled with non-interacting radiation and dust fluids is obtained.  
In Sec.~\ref{classical}, classical FLRW cosmology in HL theory is considered and analytical solutions for the scale factor are presented for open and closed universes. Some particular cases are also shown.
These solutions contains all parameters of HL theory except  the cosmological constant-like term, which is neglected in this work.
In Sec.~\ref{quantumcosmology}, I have obtained non-singular solutions for quantum FLRW cosmology in HL theory, which are the results of this paper. 
From the canonical quantization of the Hamiltonian, a Wheeler-DeWitt equation was obtained and analytically solved.
From the solutions for the wave function, using the dBB interpretation of quantum mechanics I have obtained analytical solutions for the scale factor for non-interacting radiation and dust fluids for open and closed universes.
Single (contracting/expanding) bounce universe solutions  were obtained as well as (multiple bounce) cyclic ones.  
The analytical quantum potential was also derived for these solutions, which is the responsible for the occurrence of non-singular behavior in the scale factor evolution.
In the Appendix~\ref{appendix}, the motivation for the initial condition used in Sec.~\ref{quantumcosmology} is discussed.
Finally, the concluding remarks and future perspectives are presented in Sec.~\ref{conclusions}.
Throughout this work I am using the natural units system, in which
$c=\hbar=1$.

\section{FLRW Cosmology in H\v{o}rava–Lifshitz Gravity}
\label{FLRWcosmology}

The FLRW metric for a homogeneous and isotropic space-time is written as:
\begin{eqnarray}\label{FLRW}
ds^2=-N(t)dt^2+a(t)^2\left(\frac{dr^2}{1-kr^2}+r^2d\Omega^2\right),
\end{eqnarray}
where $N(t)$ is the lapse function, $a(t)$ is the scale factor, $d\Omega^2$ is the line element of a 2-sphere with unitary radius and $k$ is the curvature constant of spatial sections, which is $k = 1$, $0$ and $-1$ for closed, flat and open universes, respectively~\footnote{I will not discuss about the possibility of the Universe being open or closed. However, Refs.~\cite{Renevey:2020zdj,Handley:2019tkm} provide more details about the possibility of a closed Universe.}.
The 4-metric is defined as $g_{\mu\nu}$, where Greek indexes run from $0$ to $3$, and  spatial 3-metric is defined as $h_{ij}$, where the Latin indexes run from $1$ to $3$.

In the following subsections, I introduce the Hamiltonian densities for the gravitational and matter sections in HL theory.

\subsection{Gravitational Hamiltonian}

In the HL theory, an
anisotropic scaling between space and time is introduced as:
\begin{eqnarray}\label{bz}
t\rightarrow b^z t\quad , \quad \vec{x}\rightarrow b \vec{x}, 
\end{eqnarray}
where $b$ is a scale parameter and $z$ is a dynamical critical exponent.
While the UV sector requires $z=3$, which breaks Lorentz invariance, in the IR sector it is recovered for $z=1$ (see Refs.~\cite{Horava:2009uw,Horava:2009if} for details). 

The gravitational action of HL theory is composed of kinetic and potential parts. 
The former generalizes the Einstein-Hilbert action and is given in terms of the extrinsic curvature (and derivatives) and a free parameter $\lambda$, which reduces to GR kinetic term in the limit  $\lambda\rightarrow 1$.
On the other hand, the potential part depends only on the 3-metric (and spatial derivatives). The detailed balance condition refers to the choice of potential part, which simplifies the theory by reducing the number of terms and facilitating its renormalization. However, the authors of Refs.~\cite{Sotiriou:2009bx,Sotiriou:2009gy} show that one should avoid this condition. 

Following the lines of  Refs.~\cite{Vakili:2013wc,Oliveira-Neto:2017yui,Oliveira-Neto:2019tvo}, in this paper I consider the projectable HL gravity without the detailed balance condition,
for $z = 3$ in $(3 + 1)$ dimensions~\cite{Bertolami:2011ka}, whose action reads:
\begin{eqnarray}\label{SHL}
S_{\rm HL}&=&\frac{M_{\rm Pl}^2}{2}\int\limits_{\mathcal{M}} d^3x\, dt N \sqrt{h}
\left[
K_{ij}K^{ij}-\lambda K^2 
- g_0 M_{\rm Pl}^2-g_1 R - M_{\rm Pl}^{-2}\left(g_2 R^2+g_3 R_{ij}R^{ij}\right)+ \right.\nonumber\\
&-&\left. M_{\rm Pl}^{-4}\left(g_4 R^3 + g_5 R R^{i}_j R^j_i + g_6 R^i_j R^j_k R^k_i + g_7 R \nabla^2 R +g_8 \nabla_i R_{jk}\nabla^i R^{jk}\right)
\right]+M_{\rm Pl}^{2}\int\limits_{\partial \mathcal{M}} d^3x \sqrt{h} K,
\end{eqnarray}
where $M_{\rm Pl}=1/\sqrt{8\pi G}$ is the Planck mass,
$h_{ij}$ is defined on the
boundary $\partial \mathcal{M}$ of the 4-dimensional manifold $\mathcal{M}$,
$h$ is the determinant of $h_{ij}$, 
$K_{ij}$ is extrinsic curvature tensor,
$K$ is the trace of $K_{ij}$,
$R_{ij}$ is the Ricci tensor,
$R$ is the Ricci scalar and $\lambda$ and $g_i$ ($i=0,..,8$) are HL parameters involved in corrections to GR.
 
In the case of FLRW space-time, the metric and the action are given by Eqs.~\eqref{FLRW} and~\eqref{SHL}, respectively. From the results of Ref.~\cite{Vakili:2013wc}, the  Lagrangian density in HL theory reads:
\begin{eqnarray}\label{LHL}
\!\!\!\!\!\!\!\!\!\!\!\mathcal{L}_{\rm HL}
=
N\left(-\frac{a\dot{a}^2}{N^2}+g_c k a -g_\Lambda a^3 - \frac{g_r k^2}{a} - \frac{g_s k}{a^3}\right),
\end{eqnarray}
where the HL parameters are defined as~\cite{Maeda:2010ke}: 
\begin{align}\label{gHL}
g_c=\frac{2}{3\lambda-1}\quad,\quad  g_\Lambda=\frac{2\Lambda}{3(3\lambda-1)}\quad,\quad  
g_r=\frac{4(3g_2+g_3)}{(3\lambda-1)M_{\rm Pl}^2}\quad,\quad g_s=\frac{8(9g_4+3g_5+g_6)}{(3\lambda-1)M_{\rm Pl}^4},
\end{align}
and the parameter $g_c$ is positive definite.
The subscripts refer to the fluid-like behavior of each term: $(g_c,\, g_\Lambda,\, g_r,\, g_s)=$ (curvature, cosmological constant,  radiation, stiff matter)-like term.

We note that the scale factor is the only variable in Eq.~\eqref{LHL}. 
In order to obtain the Hamiltonian formulation, we must compute $P_a$, the momentum canonically conjugated to $a$, which reads:
\begin{eqnarray}\label{Pa}
P_a=\frac{\partial \mathcal{L}_{HL}}{\partial \dot{a}}=-\frac{2 a\dot{a}}{N}.
\end{eqnarray}
Using the definition of $P_a$, the Hamiltonian density reads:
\begin{eqnarray}\label{HHL}
\!\!\!\!\!\!\!\!\!H_{\rm HL}
=
N\left(-\frac{P_a^2}{4a}-g_c k a +g_\Lambda a^3 + \frac{g_r k^2}{a} + \frac{g_s k}{a^3}\right).
\end{eqnarray}
The lapse function $N$ appears in $H_{\rm HL}$ as a lagrangian multiplier (there is no momentum canonically conjugated to $N$).
Rewriting $H_{\rm HL}$ as $H_{\rm HL}=N\mathcal{H}_{\rm HL}$, when we vary $H_{\rm HL}$ with respect to $N$, we obtain~\cite{Pinto-Neto:2013toa} that
$\mathcal{H}_{\rm HL}\approx 0$ ($\approx$ means weakly zero). This is called the {\it super-Hamiltonian constraint}.

\subsection{Matter Hamiltonian}

A perfect fluid can be described by a Hamiltonian using the Schutz formalism~\cite{Schutz}. 
The action for a perfect fluid in this formalism reads:
\begin{eqnarray}\label{Sm}
S_{\rm m}=\int d^4x \sqrt{-g} \,(16\pi p),
\end{eqnarray}
where $g$ is the determinant of $g_{\mu\nu}$ and $p$ is the pressure of the fluid.
The equation of state for a perfect fluid is written as:
\begin{eqnarray}\label{barotropic}
p=\omega \rho,
\end{eqnarray}
where $\rho$ is the energy density and $\omega$ is the equation of state parameter, which is subject to $-1\leq\omega\leq 1$ . 

The Schutz formalism consists of writing the fluid 4-velocity $U_\mu$ in terms of six velocity-potentials, which in FRLW cosmology these potentials are reduced to three.
Substituting the FLRW metric~\eqref{FLRW} into the $S_m$ action~\eqref{Sm}, 
identifying  the canonical variables and performing some canonical transformations~\cite{Lapchinsky:1977vb}, we obtain a simple Hamiltonian for a single perfect fluid, which reads:
\begin{eqnarray}\label{Hm}
H_{\rm m}
=
N\frac{P_T}{a^{3\omega}},
\end{eqnarray}
where $P_T$ is the momentum canonically conjugated to the fluid degree of
freedom $T$, which can be interpreted as a time variable.
The Hamiltonian $H_{\rm m}$ can also be written as $H_{\rm m}=N\mathcal{H}_{\rm m}$, where $\mathcal{H}_{\rm m}$ is a lagrangian multiplier, similarly to $\mathcal{H}_{\rm HL}$ of the gravitational part.   

\subsection{Full Hamiltonian}

From the gravitational and matter Hamiltonians, Eqs.~\eqref{HHL} and~\eqref{Hm}, respectively, the minisuperspace Hamiltonian for a single perfect fluid in HL theory reads:
\begin{eqnarray}\label{Hfull}
H
&=&
N\left(-\frac{P_a^2}{4a}-g_c k a +g_\Lambda a^3 + \frac{g_r k^2}{a} + \frac{g_s k}{a^3}+\frac{P_T}{a^{3\omega}}\right),
\end{eqnarray}
from which we can also define the super-Hamiltonian constraint $\mathcal{H}=\mathcal{H}_{\rm HL}+\mathcal{H}_{m}$, which satisfies: 
\begin{eqnarray}\label{superH}
\mathcal{H}\approx 0.
\end{eqnarray}
In the following sections, I will show that this constraint leads to the Friedmann equation in the classical level, whereas at the quantum level it gives the Wheeler-DeWitt equation.
In the latter, the super-Hamiltonian constraint is essential in the procedure of canonical quantization.

In this work, I will consider a HL quantum FLRW universe filled with non-interacting radiation ($\omega=1/3$) and dust ($\omega=0$) fluids.
However, Eq.~\eqref{Hfull} is valid only for a single fluid.
I will set $\omega=1/3$ in order to describe a radiation fluid and,
following Ref.~\cite{PintoNeto:2005gx},
I will also 
include dust as second decoupled fluid
by introducing another term of the type~\eqref{Hm} for $\omega=0$.
From these considerations, the Hamiltonian for radiation and dust fluids in HL theory reads:  
\begin{eqnarray}\label{Hraddust}
\!\!\!\!\!\!\!\!\!H_{\rm rd}
&=&
N\mathcal{H}_{\rm rd}=N\left(-\frac{P_a^2}{4a}-g_c k a +g_\Lambda a^3+ \frac{g_r k^2}{a} + \frac{g_s k}{a^3}+\frac{P_T}{a}+P_\varphi\right),
\end{eqnarray}
where $P_\varphi$ is a constant stemming from Eq.~\eqref{Hm} for $\omega=0$ and $\mathcal{H}_{\rm rd}$ is the super-Hamiltonian related to $H_{\rm rd}$.

Additionally,
for the potential part of HL theory,
I will consider from now on the particular case of no cosmological constant-like term ($g_\Lambda=0$).


\section{Classical dynamics H\v{o}rava–Lifshitz Cosmology}
\label{classical}

In this section, I consider analytical solutions of the HL classical cosmology in minisuperspace for non-interacting radiation and dust fluids. 
The equations of motion for each system variable and its canonically conjugate momentum are then obtained from the evaluation of the Poisson brackets of each of them with the Hamiltonian.

The HL super-Hamiltonian constraint containing radiation and dust fluids, Eq.~\eqref{Hraddust}, together with the equation for $P_a$,  Eq.~\eqref{Pa}, result in the classical Friedmann equation:
\begin{eqnarray}\label{Friedamnraddust}
\!\!\!\!\!\!\!\!\!\left(\frac{\dot{a}}{a}\right)^2
=
N^2\left(
\frac{g_s k}{a^6}
+ \frac{g_r k^2}{a^4} 
-\frac{g_c k}{a^2} 
+\frac{P_T}{a^4}
+\frac{P_\varphi}{a^3}
\right).
\end{eqnarray}
From the Hamiltonian, Eq.~\eqref{Hraddust}, the equations of motion for $T$ and $\varphi$ read:
\begin{eqnarray}
\dot{T}&=&\{T,H_{\rm rd}\}=\frac{N}{a}\quad \rightarrow \quad a\, dT =N\, dt,\nonumber\\
\dot{\varphi}&=&\{\varphi,H_{\rm rd}\}=N\quad\, \rightarrow \quad \ \ d\varphi=N\,dt.\nonumber
\end{eqnarray}
One needs to choose a gauge, which corresponds to choosing the lapse function $N$ in order to define the time variable. In principle, both $T$ and $\varphi$ can be the time variable.
If we choose $N=1$, $\varphi$ is the cosmic time, whereas for $N=a$ we obtain that $T$ is the conformal time.
Analytical solutions of the Friedmann equation, Eq.~\eqref{Friedamnraddust}, can only be obtained for the latter, so I will consider the  gauge $N=a$ throughout this paper, which means that $T=\eta$, where $\eta$ is the conformal time. 
Therefore, Eq.~\eqref{Friedamnraddust} now reads:
\begin{eqnarray}\label{Friedamnraddusteta}
\!\!\!\!\!\!\left(\frac{a'}{a^2}\right)^2
=
\frac{g_s k}{a^6}
+ \frac{g_r k^2}{a^4} 
-\frac{g_c k}{a^2} 
+\frac{P_\eta}{a^4}
+\frac{P_\varphi}{a^3},
\end{eqnarray}
where the prime denotes derivative with respect to conformal time.

The analytical solutions for Eq.~\eqref{Friedamnraddusteta} with the initial condition $a(0)=0$ read:
\begin{widetext}
\begin{eqnarray}\label{aclfull}
a(\eta) = \left\{ \begin{array}{lllll}
\sqrt{
\sqrt{\frac{g_s}{g_c}}\sin \left(2\sqrt{g_c} \eta\right)
+
\frac{(P_\eta+g_r)}{g_c}\sin^2 \left(\sqrt{g_c} \eta\right)
}
&+&\frac{P_\varphi }{2 g_c}\left[1-\cos \left(\sqrt{g_c} \eta\right)\right]
&, & k=1,\\
\sqrt{
\sqrt{\frac{|g_s|}{g_c}}\sinh \left(2\sqrt{g_c} \eta\right)
+
\frac{(P_\eta+g_r)}{g_c}\sinh^2 \left(\sqrt{g_c} \eta\right)
}
&+&\frac{P_\varphi }{2 g_c}\left[\cosh \left(\sqrt{g_c} \eta\right)-1\right]
&, & k=-1,
\end{array} 
\right. 
\end{eqnarray}
\end{widetext}
where $g_c$ and the product $g_s k$ are both positive definite. 
The latter is positive when $g_s>0$ ($g_s<0$) for $k= 1$ ($k=-1$). 
From now on the notation $g_s k=|g_s|$ represents the case $g_s<0,\, k=-1$ or both $g_s>0,\, k=1$ and $g_s<0,\, k=-1$ cases when the results both for $k=\pm1$ can be written in a single expression.  
From Eq.~\eqref{Friedamnraddusteta}, the stiff matter-like term involving $g_s$ initially dominates near the singularity at $\eta=0$,
followed by radiation dominance, which consists of a radiation fluid plus a ``HL radiation" term involving $g_r$.
Far from the singularity, radiation domination is followed by dust domination, and ending up with domination of the curvature term involving $g_c$. If the cosmological constant term were not neglected, as in the Hamiltonian given by Eq.~\eqref{Hraddust}, far from the singularity it would dominate after the curvature term.    

Now I consider some particular cases.
In the limit where the curvature-like term is negligible ($g_c\to0$),
Eqs.~\eqref{aclfull} read:
\begin{eqnarray}\label{aclnogc}
a(\eta)=\sqrt{2 \sqrt{|g_s|}\,\eta
+ \left( P_\eta+g_r\right)\eta^2}
+\frac{P_\varphi \eta^2}{4},
\end{eqnarray}
which is valid for both $k=\pm1$.
Also, when we additionally consider that the
stiff matter-like time is negligible ($g_s\to0$),
the latter result reduces to:
\begin{eqnarray}
a(\eta)=\sqrt{P_\eta+g_r k^2}\,\eta+\frac{P_\varphi }{4}\eta^2.
\end{eqnarray}
The remaining term from  HL theory is the radiation-like constant, $g_r$, which adds up to usual radiation.
When $k=0$ or in the limit $g_r\to0$, HL reduces to GR and one obtains a flat universe filled with radiation and dust fluids, where radiation dominates near the singularity and dust dominates far from it.

On the other hand, when stiff matter like time is negligible ($g_s\to0$) in Eqs.~\eqref{aclfull}, one obtains:
\begin{widetext}
\begin{eqnarray}\label{aclnogs}
a(\eta) = \left\{ \begin{array}{lll}
\sqrt{
\frac{(P_\eta+g_r)}{g_c}
}
\sin\left(\sqrt{g_c} \eta\right)
+\frac{P_\varphi }{2 g_c}\left[1-\cos \left(\sqrt{g_c} \eta\right)\right]
&, & k=1,\\
\sqrt{
\frac{(P_\eta+g_r)}{g_c}
}
\sinh \left(\sqrt{g_c} \eta\right)
+\frac{P_\varphi }{2 g_c}\left[\cosh \left(\sqrt{g_c} \eta\right)-1\right]
&, & k=-1.
\end{array} 
\right.  
\end{eqnarray}
\end{widetext}

In the following I will consider the HL quantum solutions.


\section{Quantum dynamics H\v{o}rava–Lifshitz Cosmology}
\label{quantumcosmology}

In this section, I consider analytical solutions of the HL quantum cosmology in minisuperspace for non-interacting radiation and dust fluids. 
The cyclic and bounce universe solutions presented in this section are obtained  
for the first time for HL theory for nonzero $g_c,\,g_r,\,g_s$ and for radiation and dust fluids.

The HL super-Hamiltonian constraint containing radiation and dust fluids, from Eq.~\eqref{Hraddust}, reads:
\begin{eqnarray}\label{HraddustQU}
\!\!\!\!\!\!\!\!\!\!\!\!\!\!\!&&
\mathcal{H}_{rd}=-\frac{P_a^2}{4a}-g_c k a + \frac{g_r k^2}{a} + \frac{g_s k}{a^3}+\frac{P_\eta}{a}+P_\varphi\approx 0,
\end{eqnarray}
where I have considered the lapse function $N=a$ and $g_\Lambda=0$ as in the Sec.~\ref{classical}. Using the Dirac formalism for constrained systems, the super-Hamiltonian
constraint is promoted to
an operator, which annihilates the quantum wave
function of the universe, $\Psi$, in the form:
\begin{eqnarray}\label{WDW}
\hat{\mathcal{H}}_{\rm rd}\Psi=0.
\end{eqnarray}
This is the so called Wheeler-DeWitt equation. From Eqs.~\eqref{HraddustQU} and~\eqref{WDW}, one obtains:
\begin{eqnarray}\label{WDWraddust}
\!\!\!\!\!\!\!\!\!
i\partial_\eta \Psi
=
\left(\frac{1}{4}\partial_a^2-g_c k a^2 + g_r k^2 + \frac{g_s k}{a^2}-i\,a\,\partial_\varphi \right)\Psi,
\end{eqnarray}
where $\hat{P}_a=-i\partial_a$, $\hat{P}_\eta=-i \partial_\eta$ and $\hat{P}_\varphi=-i \partial_\varphi$~(see Refs.~\cite{PintoNeto:2005gx,Lapchinsky:1977vb}) and $\Psi=\Psi(a,\varphi,\eta)$.
However, two important considerations must be made. Firstly, there is an operator-ordering ambiguity in the Wheeler-DeWitt equation, which is related to the choice of measure in the path integral in the canonical quantization procedure when one replaces
the momentum $P_a$ by its corresponding operator~\cite{Halliwell:1988wc}.
In order to account for this ambiguity, the kinetic term must be rewritten properly.
Secondly, in order to get rid of the derivative $\partial_\varphi$, following Ref.~\cite{PintoNeto:2005gx} I will consider that the wave function $\Psi$ is an eigenstate of the dust matter operator, such that $\hat{P}_\varphi |\Psi\rangle =P_\varphi|\Psi\rangle$, which implies that dust matter is conserved~\footnote{As a first attempt, I consider this simple case, where the evolution is non-unitary. However, one can manage to obtain an unitary solution in which dust matter creation is possible ~\cite{PintoNeto:2005gx}.}.
Therefore, the wave function $\Psi$ in an eigenstate of $\hat{P}_\varphi$ with eigenvalue $P_\varphi$ and
the wave function $\Psi=\Psi(a,\varphi,\eta)$ can be written as:
\begin{eqnarray}\label{psidust}
\Psi(a,\varphi,\eta)=\Psi(a,\eta)e^{i P_\varphi \varphi}.
\end{eqnarray}
From these considerations, Eq.~\eqref{WDWraddust} now reads:
\begin{eqnarray}\label{WDWraddustalpha}
\!\!\!\!\!\!\!\!\!
i\partial_\eta \Psi(a,\eta)
&=&
\left(
-\frac{1}{4}\partial_a^2 
+ \frac{\alpha}{4a}\partial_a+ g_c k a^2 
- g_r k^2 
- \frac{g_s k}{a^2}
-P_\varphi\,a \right)\Psi(a,\eta),
\end{eqnarray}
where the parameter $\alpha$ represents the ambiguity in the ordering of $a$ and $P_a$ in the kinetic term
of Eq.~\eqref{HraddustQU}
and the transformation $\eta\to-\eta$ was also considered.
The appropriate choice of $\alpha$ will be useful in the following calculations, although the results must not depend on it. 
On the other hand, the sign change in the time variable was done in
order to write the Wheeler-DeWitt equation as a Schrödinger-type equation (except for the term involving $\alpha$, at this point).

\subsection{De Broglie-Bohm Interpretation}\label{dBB}

The wave function $\Psi$ can be written in the polar form as $\Psi=R\, e^{i\, S}$, such that imaginary and real parts give evolution equations for $R$ and $S$, respectively,
which are real functions. 
In order to introduce the dBB interpretation, I write the Lagrangian for Eq.~\eqref{WDWraddustalpha}, which reads:
\begin{eqnarray}\label{Lagrangian}
\mathcal{L} _ {\rm rd} &=& 
 a^{-\alpha}\left[
i\Psi^*\partial_\eta\Psi 
- \frac{1}{4}\partial_a \Psi^*\partial_a\Psi 
- V\Psi^*\Psi  \right],
\end{eqnarray}
where 
\begin{eqnarray}\label{V}
V=g_c k a^2 - \frac{g_s k}{a^2} - g_r k^2-P_\varphi a,
\end{eqnarray}
is an external classical potential.
From Noether's theorem~\cite{PESKIN}, invariance of $\Psi$ under internal symmetry ($\Psi\to e^{i \theta}\Psi$)  results in a conserved charge $\rho$ and a conserved current $J$, which are related by the following continuity equation:
\begin{eqnarray}\label{R}
\partial_\eta\left(a^{-\alpha} R^2\right)+\partial_a\left(a^{-\alpha}R^2\frac{\partial_a S}{2}\right)=0,
\end{eqnarray}
where $\rho=a^{-\alpha} R^2$ and $J=a^{-\alpha}R^2(\partial_a S)/2$. This equation is exactly the imaginary part of Eq.~\eqref{WDWraddustalpha} mentioned before and is interpreted as the equation of conservation of probability.

In the dBB interpretation, particles have a deterministic trajectory, which is given by $a(\eta )$ in the present case.
Therefore, an equation of motion must be postulated. This equation can be built out of $J=\rho\, v$, where $v=a'$ is the velocity of the particle. Therefore, 
\begin{eqnarray}\label{v}
a'=\frac{\partial_a S}{2},
\end{eqnarray}
which is known as the {\it guidance equation}~\footnote{For arbitrary lapse function $N$ and equation of state parameter $\omega$, and  undoing the transformation $\eta\to-\eta$, $a'=-\frac{N}{2a}(\partial_a S)$.}. From the knowledge of $S$ and an initial condition for $a$, one can integrate it to obtain $a(\eta)$. 
Also, one can notice the trajectory of the particle is independent of the choice of $\alpha$.

On the other hand, the real part of Eq.~\eqref{WDWraddustalpha} in terms of $R$ and $S$ reads:
\begin{eqnarray}\label{S}
\partial_\eta S +\frac{(\partial_a S)^2}{4}
+V
+Q
=0.
\end{eqnarray}
This equation is a Hamilton–Jacobi-type equation, where the last term, 
\begin{eqnarray}\label{Q}
Q=-\frac{1}{4}\frac{\partial_a^2(a^{-\alpha}R)}{(a^{-\alpha}R)},
\end{eqnarray}
is the so called {\it quantum potential}.
Therefore, from Eq.~\eqref{S},
one concludes that deterministic trajectories are subject to classical and quantum potentials. 

Therefore, in order to obtain a solution for $a(\eta)$ from Eq.~\eqref{v}, one needs to solve both Eqs.~\eqref{R} and~\eqref{S}, which are coupled equations for $R$ and $S$.

\subsection{Analytical Results}

The main goal of this paper is to present analytical solutions for Eq.~\eqref{WDWraddustalpha}. 
There is a solution
for a similar problem in the literature, which can be adapted for the present case. 
However, some changes of variables must be done in order to use this solution. These details will be presented in the following.
From the analytical solutions of the Wheeler-DeWitt equation, analytical solutions can be obtained for the scale factor, solving Eq.~\eqref{v}. 

The aforementioned solution is given in Ref.~\cite{PintoNeto:2005gx}. 
This solution was given for the particular case of Eq.~\eqref{WDWraddustalpha}
when $g_c=1$, $g_r=0$ and $g_s=0$. Additionally,
the authors set $\alpha=0$.
The Wheeler-DeWitt equation for this case (Eq.~(25) from Ref.~\cite{PintoNeto:2005gx}) reads:
\begin{eqnarray}\label{schoringer}
\!\!\!\!\!\!\!\!\!\!\!\!\!\!
i\partial_\eta \psi(a,\eta)
&=&
\left(
-\frac{1}{2m}\partial_a^2 
+ \frac{m\omega_0^2}{2} a^2 
- P_\varphi\,a \right)\psi(a,\eta).
\end{eqnarray}
This is a one-dimensional Schrödinger equation for a particle with mass $m$ and a potential which contains an harmonic oscillator term with frequency $\omega_0$ and a constant force $P_\varphi$.
In order to use this solution for the present case, one needs to reduce Eq.~\eqref{WDWraddustalpha} to Eq.~\eqref{schoringer} performing some changes of variables. One also needs to set $\alpha\neq0$, which is in fact the key point here and, as shown in Sec.~\ref{dBB}, it does not affect the result for the scale factor.

I consider the following change of variables for the wave function of Eq.~\eqref{WDWraddustalpha}:
\begin{eqnarray}\label{psidustgr}
\Psi(a,\eta)
=
e^{ig_rk^2\eta}
a^{\alpha/2}
\mu(a,\eta),
\end{eqnarray}
where $\alpha=-1\pm\sqrt{1+16 g_s k}$.
Both values of $\alpha$ are suitable to absorb the term $g_s/a^2$. 
I choose:
\begin{eqnarray}\label{alpha}
\alpha=-1+\sqrt{1+16 g_s k},
\end{eqnarray}
where one obtains $\alpha\to 0$ when $g_s\to 0$.
Also, the term $-g_r k^2$ is also absorbed in the complex exponential.
From these change of variables, Eq.~\eqref{WDWraddustalpha} is now an equation for $\mu(a,\eta)$, which reads:
\begin{eqnarray}\label{WDWraddustgseq0mu}
\!\!\!\!\!\!\!\!\!
i\partial_\eta \mu(a,\eta)
&=&
\left(
-\frac{1}{4}\partial_a^2 
+ g_c k a^2 
- P_\varphi\,a \right)\mu(a,\eta).
\end{eqnarray}
This equation is exactly Eq.~\eqref{schoringer} for the particular case when:
\begin{eqnarray}
\label{m}
\quad m&=&2,\\
\label{omega}
\omega_0&=&\sqrt{g_c\,k}.
\end{eqnarray}
For these values of $\omega_0$ and $m$, the analytical result for the wave function, given by Eq.~(41) of Ref.~\cite{PintoNeto:2005gx}, reads:
\begin{align}
\mu(a,\eta)
&=\left(\frac{8\bar{\sigma}}{\pi}\right)^{1/4}\sqrt{
\frac{1}{\cos\left(\sqrt{g_ck}\eta\right)\left[1+i\bar{\sigma}\frac{\tan\left(\sqrt{g_ck}\eta\right)}{\sqrt{g_ck}}\right]}
} 
\exp{
\left\{
\frac{i\sqrt{g_ck}}{\tan\left(\sqrt{g_ck}\eta\right)}
\right.}\notag\\
&
\times\left.\left[
a^2
-
\frac{\left(a-\frac{P_\varphi}{2}\frac{\left[1-\cos\left(\sqrt{g_ck}\eta\right)\right]}{g_c k}\right)^2}{\cos\left(\sqrt{g_ck}\eta\right)^2\left[1+i\bar{\sigma}\frac{\tan\left(\sqrt{g_ck}\eta\right)}{\sqrt{g_ck}}\right]}
+aP_\varphi\frac{1-\cos\left(\sqrt{g_ck}\eta\right)}{g_c k\cos\left(\sqrt{g_ck}\eta\right)}
\right.\right.\notag\\
&\!\!\!\left.\left.
+
\frac{P_\varphi^2}{2g_ck}\left(\eta\frac{\tan\left(\sqrt{g_ck}\eta\right)}{\sqrt{g_ck}}-\frac{1-\cos\left(\sqrt{g_ck}\eta\right)}{g_ck\cos\left(\sqrt{g_ck}\eta\right)}\right)
\right]
\right\}.
\label{psiref}
\end{align}
The solution for $\mu(a,\eta)$ was obtained using the propagator of the forced quantum harmonic oscillator using the following initial state: 
\begin{eqnarray}\label{psi0}
\mu_0(a,0)=\left(\frac{8\, \bar{\sigma}}{\pi}\right)^{1/4}e^{-\bar{\sigma} a^2},
\end{eqnarray}
where $\bar{\sigma}=\sqrt{\sigma^2-g_s k}+i\sqrt{g_s k}$.
One must notice that both the solution and its initial condition contain the parameter $g_s$ in the constant $\bar{\sigma}$, although there is no term involving $g_s$ in Eq.~\eqref{WDWraddustgseq0mu}.
In fact, only the parameters $\sigma$ and $P_\varphi$, which are related to radiation and dust fluids, respectively, were expected. 
Additionally, the dynamics for this matter content is radiation domination near $\eta=0$ and dust domination far from it.
Therefore, an initial condition at $\eta=0$ must involve only the parameter $\sigma$.
However, the full problem considered in this section before the changes of variables involves a $g_s$ term, which, from the classical dynamics of Sec.~\ref{classical}, behaves as a stiff matter-like fluid and dominates over radiation near $\eta=0$.
Then, one should expect that an initial condition for this problem involves uniquely $g_s$.
Nonetheless, due to the fact that the $g_s$ term does not represent a fluid, radiation fluid degrees of freedom must also be present in the initial condition, i.e., it must be a function of both $g_s$ and $\sigma$.
In other words, although changes of variables are performed and $g_s$ is no longer explicit in Eq.~\eqref{WDWraddustgseq0mu}, the $g_s$ information is encoded in $\bar{\sigma}$ due to its relevance near $\eta=0$. 
In the Appendix~\ref{appendix}, I derive an ansatz solution for the particular case of a radiation dominated quantum FLRW universe in the framework of HL theory where only the $g_s$ parameter is nonzero. The obtained ansatz justifies this choice of initial condition. 
Analogously, the $g_r$ radiation fluid-like term is encoded in radiation fluid degree of freedom, $\sigma$.

From this analytical result, the expression for $\Psi(a,\varphi,\eta)$, from Eqs.~\eqref{psidust},~\eqref{psidustgr} and~\eqref{alpha}, reads:
\begin{eqnarray}\label{Psigs0}
\!\!\!\!\!\!\!\!\!\!\!\!\!\!\!
\Psi(a,\varphi,\eta)
=
e^{ig_rk^2\eta}
e^{i P_\varphi \varphi}
a^{\left(-1+\sqrt{1+16 g_s k}\right)/2}
\mu(a,\eta),
\end{eqnarray}
where $\mu(a,\eta)$ is given by the solution of Eq.~\eqref{psiref},

From Eq.~\eqref{Psigs0}, one can obtain $S(a,\eta)$ and substitute it in Eq.~\eqref{v}. Solving this equation for the initial condition $a(0)=a_B$, one obtains:
\begin{widetext}
\begin{eqnarray}\label{aqu}
a(\eta) = \left\{ \begin{array}{lll}
a_B \sqrt{
\cos^2 \left(\sqrt{g_c} \eta\right)
+
\sqrt{\frac{g_s}{g_c}}\sin \left(2\sqrt{g_c} |\eta|\right)
+\frac{\sigma^2}{g_c}\sin^2 \left(\sqrt{g_c} \eta\right)
}
+\frac{P_\varphi }{2 g_c}\left[1-\cos \left(\sqrt{g_c} \eta\right)\right]
&, & k=1,\\
a_B \sqrt{\cosh^2 \left(\sqrt{g_c} \eta\right)
+
\sqrt{\frac{|g_s|}{g_c}}\sinh \left(2\sqrt{g_c} |\eta|\right)
+\frac{\sigma^2}{g_c}\sinh^2 \left(\sqrt{g_c} \eta\right)
}
+\frac{P_\varphi }{2 g_c}\left[\cosh \left(\sqrt{g_c} \eta\right)-1\right]
&, & k=-1.
\end{array} 
\right. 
\end{eqnarray}
\end{widetext}
These are quantum bounce solutions, which, compared to the classical ones, the squared trigonometric and hyperbolic cosine novel terms play the role of avoiding singularities.
These results enable the study of the background quantum cosmology for this model from an analytical perspective.

The positive spatial section solution represents a cyclic universe (no big bang and big crunch singularities), whereas the negative spatial section solution has a unique bounce, which connects a contraction and an expansion phase.
However, one must consider some limits for the constants $g_c$ and $g_s$ in order to appreciate the overlap of these effects.
Some plots will be presented later in this section in order to illustrate the behavior of these solutions for different values of its constants.

Comparing Eqs.~\eqref{aqu} with the classical solutions, Eqs.~\eqref{aclfull}, 
one obtains that in the classical limit $\sigma^2=P_\eta+g_rk^2$, which means that $\sigma$ is a degree of freedom related to a radiation fluid.
Therefore, $g_r$ only contributes to an effective radiation fluid density and does not appear explicitly in $a(\eta)$.
Also, the classical behavior is obtained when 
\begin{eqnarray}\label{aqulimit}
\!\!\!\!\!\!\!\!\!\!\!\!\!\!\!\!\!\left. 
\begin{array}{rl}
&|\tan\left(\sqrt{g_c} \eta\right)|
\\
&|\tanh\left(\sqrt{g_c} \eta\right)|
\end{array} 
\right\} 
\gg
\frac{\sqrt{g_c} (\sqrt{|g_s|+\sigma^2}-\sqrt{|g_s|})}{\sigma ^2},
\end{eqnarray}
for $k=1$ and $k=-1$, respectively. 
In the limit where $g_c\to1$ and $g_s\to0$,
these results reduce to $1/\sigma$ for both values of $k$, where $\sigma$ contains the $g_rk^2$ term. Also, both results reduce to $|\eta|\gg 1/\sigma$ when one additionally considers the limit $k\to 0$ (see Ref.~\cite{PintoNeto:2005gx}).

One must draw attention to the module $|\eta|$ in the terms involving $g_s$ in Eqs.~\eqref{aqu}. 
This is because the wave function given in Eq.~\eqref{psiref} was obtained by the propagation of the initial condition only for negative values of $\eta$.
Therefore, one needs to solve also for positive values of $\eta$ (setting $\eta\to-\eta$ in both Eqs.~\eqref{v} and~\eqref{psiref}) and incorporate it in the solutions.
Due to the presence of odd trigonometric and hyperbolic sines, the module 
ensures that the solution is symmetric, which can be confirmed from the classical solutions of Sec.~\ref{classical}. 
This observation was not necessary in the calculations of Ref.~\cite{PintoNeto:2005gx}, because only even functions appear in the results for $a(\eta)$.

As in Sec.~\ref{classical}, I will consider some particular cases of Eqs.~\eqref{aqu}.
In the limit where $g_c\to0$, for both values of $k$ it reduces to:
\begin{eqnarray}\label{aqugc0}
\!\!\!\!\!\!\!\!\!\!\!
a(\eta)=a_B\sqrt{1+ 
2 \sqrt{|g_s|}\, |\eta|+ \sigma^2\eta^2 }
+\frac{P_\varphi \eta^2}{4}.
\end{eqnarray}
Also considering in the latter equation the limit where $g_s\to0$, it reduces to:
\begin{eqnarray}\label{aqugc0gs0}
\!\!\!\!\!\!\!\!\!\!\!
a(\eta)=a_B\sqrt{1+ \sigma^2\eta^2}
+\frac{P_\varphi \eta^2}{4},
\end{eqnarray}
where now $\bar{\sigma}=\sigma$.
This result resembles the case of a FLRW flat universe in GR with radiation and dust~\cite{PintoNeto:2005gx}, except for the HL radiation term correction for $k\neq0$. 
When $k=0$, HL theory reduces to GR and the results are exactly the same. 

On the other hand, when $g_s\to0$, Eqs.~\eqref{aqu} reduce to:
\begin{widetext}
\begin{eqnarray}\label{aqugs0}
a(\eta) = \left\{ \begin{array}{lll}
a_B \sqrt{
\cos^2 \left(\sqrt{g_c} \eta\right)
+\frac{\sigma^2}{g_c}\sin^2 \left(\sqrt{g_c} \eta\right)
}
+\frac{P_\varphi }{2 g_c}\left[1-\cos \left(\sqrt{g_c} \eta\right)\right]
&, & k=1,\\
a_B \sqrt{\cosh^2 \left(\sqrt{g_c} \eta\right)
+\frac{\sigma^2}{g_c}\sinh^2 \left(\sqrt{g_c} \eta\right)
}
+\frac{P_\varphi }{2 g_c}\left[\cosh \left(\sqrt{g_c} \eta\right)-1\right]
&, & k=-1.
\end{array} 
\right. 
\end{eqnarray}
These results are qualitatively similar to the ones of Ref.~\cite{PintoNeto:2005gx} when $g_c=1$, which are given by:
\begin{eqnarray}\label{aqunogsgc1}
a(\eta) = \left\{ \begin{array}{lll}
a_B \sqrt{
\cos^2 \left(\eta\right)
+\sigma^2\sin^2 \left(\eta\right)
}
+\frac{P_\varphi }{2}\left[1-\cos \left(\eta\right)\right]
&, & k=1,\\
a_B \sqrt{\cosh^2 \left( \eta\right)
+\sigma^2\sinh^2 \left( \eta\right)
}
+\frac{P_\varphi }{2}\left[\cosh \left(\eta\right)-1\right]
&, & k=-1.
\end{array} 
\right. 
\end{eqnarray}
\end{widetext}

In order to illustrate the results, I present some
plots of the scale factor for representative values, which capture the effect of each constant in its evolution.
I will consider $P_\varphi=0$ for all plots and focus on the effects of $g_c$ and $g_s$, which are introduced by HL theory. 
Additionally, in all plots there is a {\it gray curve} which corresponds to the solution with respect to GR, where the HL parameters are null, given by Eq.~\eqref{aqugc0gs0}.

In Figure~\ref{fig1}, 
the scale factor is considered in the limit where $g_s\to0$, given by Eqs.~\eqref{aqugs0}.
In panel (a), one notices that for $k=1$ nonzero values of $g_c$ produce oscillating universes, which are cyclic and with multiple bounces.
Starting from $g_c=0$, for increasing values of $g_c$ the oscillation frequency grows and the oscillation amplitude becomes smaller.
In panel (b), for $k=-1$, increasing values of $g_c$ make the bounce more curved.

In Figure~\ref{fig2}, 
the scale factor is considered in the limit where $g_c\to0$, given by Eq.~\eqref{aqugc0}.
In this case, for both values of the spatial section, the bouncing universes become more curved for increasing values of $g_s$.
This is qualitative similar to the case $g_s\to0$ for $k=-1$ of panel (b) of Fig.~\ref{fig1}, where it is the increase of $g_c$ that progressively curves the universes.
However, their difference is that in the present case $a(\eta)/a_B\approx 1+\sqrt{g_s}|\eta|+\mathcal{O}\left(\eta ^2\right)$ near the bounce, whereas $a(\eta)/a_B\approx 1+ \left(g_c+\sigma^2\right)\eta ^2/2 +\mathcal{O}\left(\eta ^3\right)$ in panel (b) of Fig.~\ref{fig1}.
Therefore, the bounce solution is more abrupt in the present case.

In Figure~\ref{fig3},
the scale factor is considered for nonzero values of both $g_c$ and $g_s$, given by Eqs.~\eqref{aqu}.
The effects are overlapped, which is the more general case.
In panel (a), where $k=1$, one notices that the oscillatory behavior due to nonzero values of $g_c$ is affected by nonzero values of $g_s$ in two aspects: (i) in the oscillatory regime, the frequency and the oscillation amplitude increase (far from the bounce) and (ii) the bounce around $\eta=0$ is linear, whereas the other bounces are qualitatively similar and deeper.  
In panel (b), where $k=-1$, the effects of both increasing values of $g_c$ and $g_s$, as I have shown in Fig.~\ref{fig1} panel (b) and Fig.~\ref{fig2}, make the bounce more curved.
Therefore, the effects add up.
However, near the bounce the scale factor reads 
$a(\eta)/a_B
\approx 
1 + \sqrt {g_s}|\eta| 
+
\left(g_c-|g_s |+\sigma^2\right)\eta^2/2 + \mathcal{O}\left (\eta^3 \right)$, which means that $g_s$ dominates near the bounce, whereas as we move away from the bounce both effects become relevant and $g_c$ eventually dominates over $g_s$.

\begin{widetext}
\begin{center}
\begin{figure}[htb!]
\subfigure[]{\includegraphics[width=7.3cm]{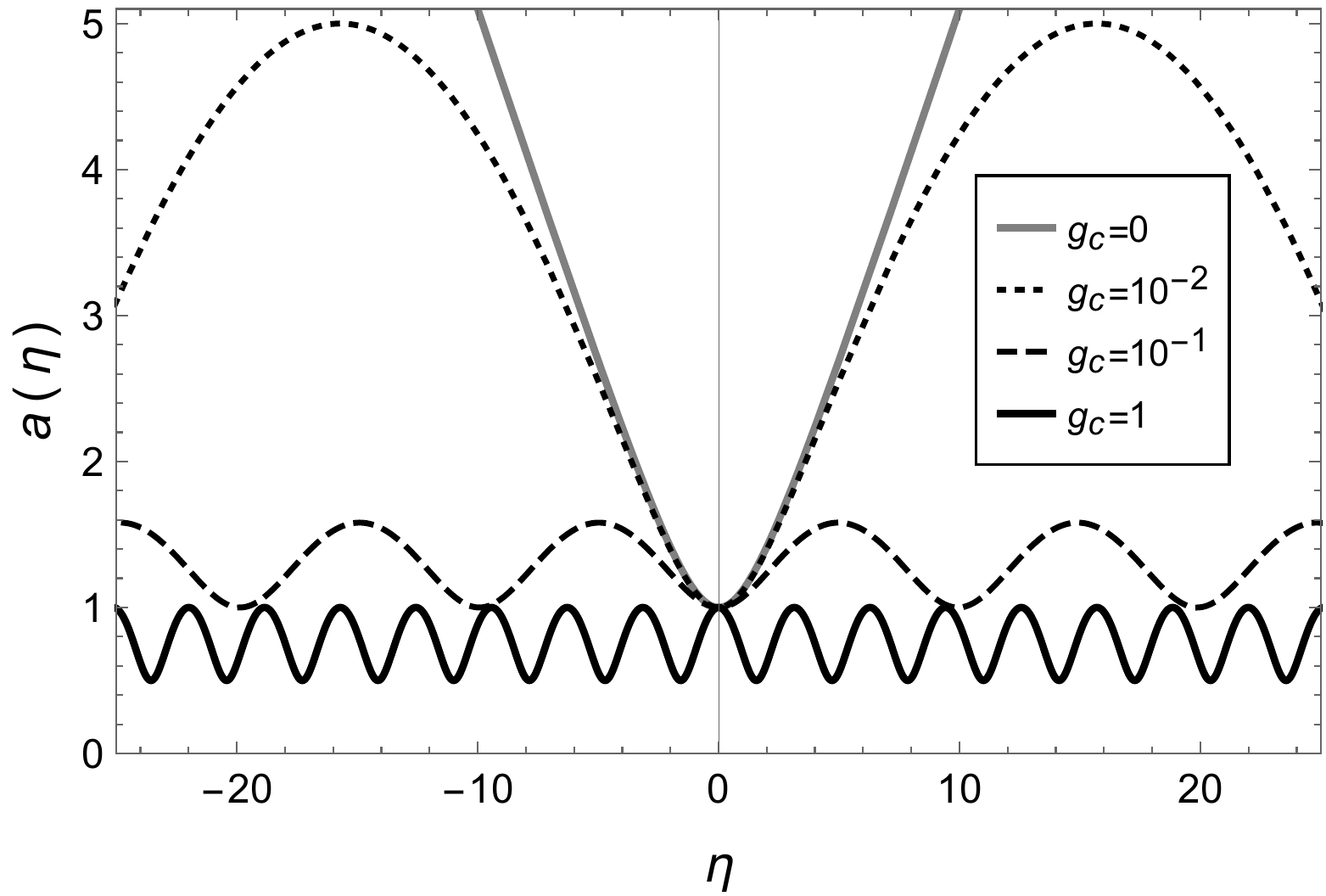}}
\subfigure[]{\includegraphics[width=7.3cm]{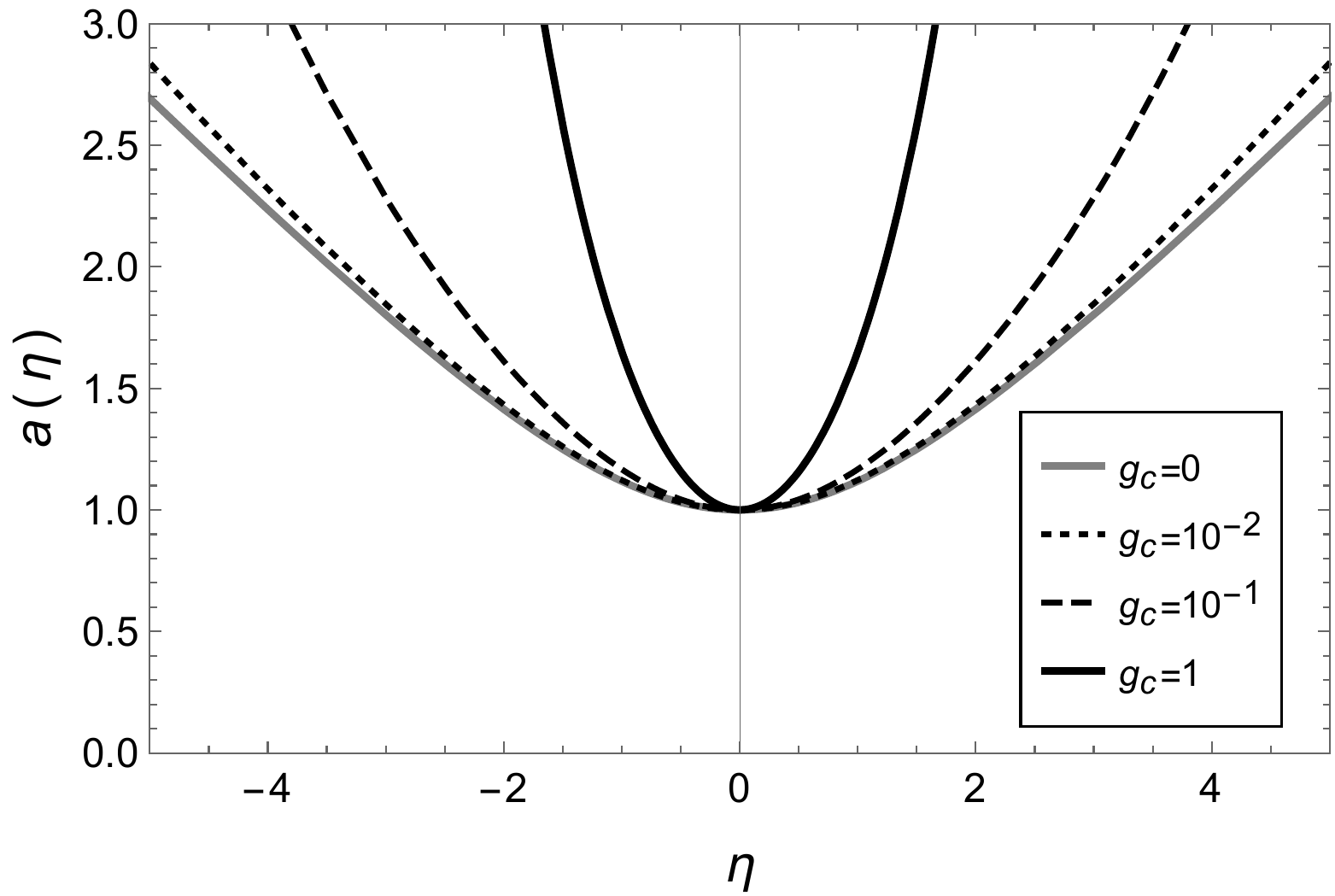}}
\caption{
Analytical scale factor in quantum HL theory as a function of the conformal time  in the limit $g_s\to0$. 
The plots are given for $a_B=1$, $P_\varphi=0$, $\sigma=0.5$ and some representative values of $g_c$.
In panels (a) and (b) the spatial sections $k=1$ and $k=-1$ are considered, respectively.}
\label{fig1}
\end{figure}
\end{center}

\begin{center}
\begin{figure}[htb!]
\includegraphics[width=7.3cm]{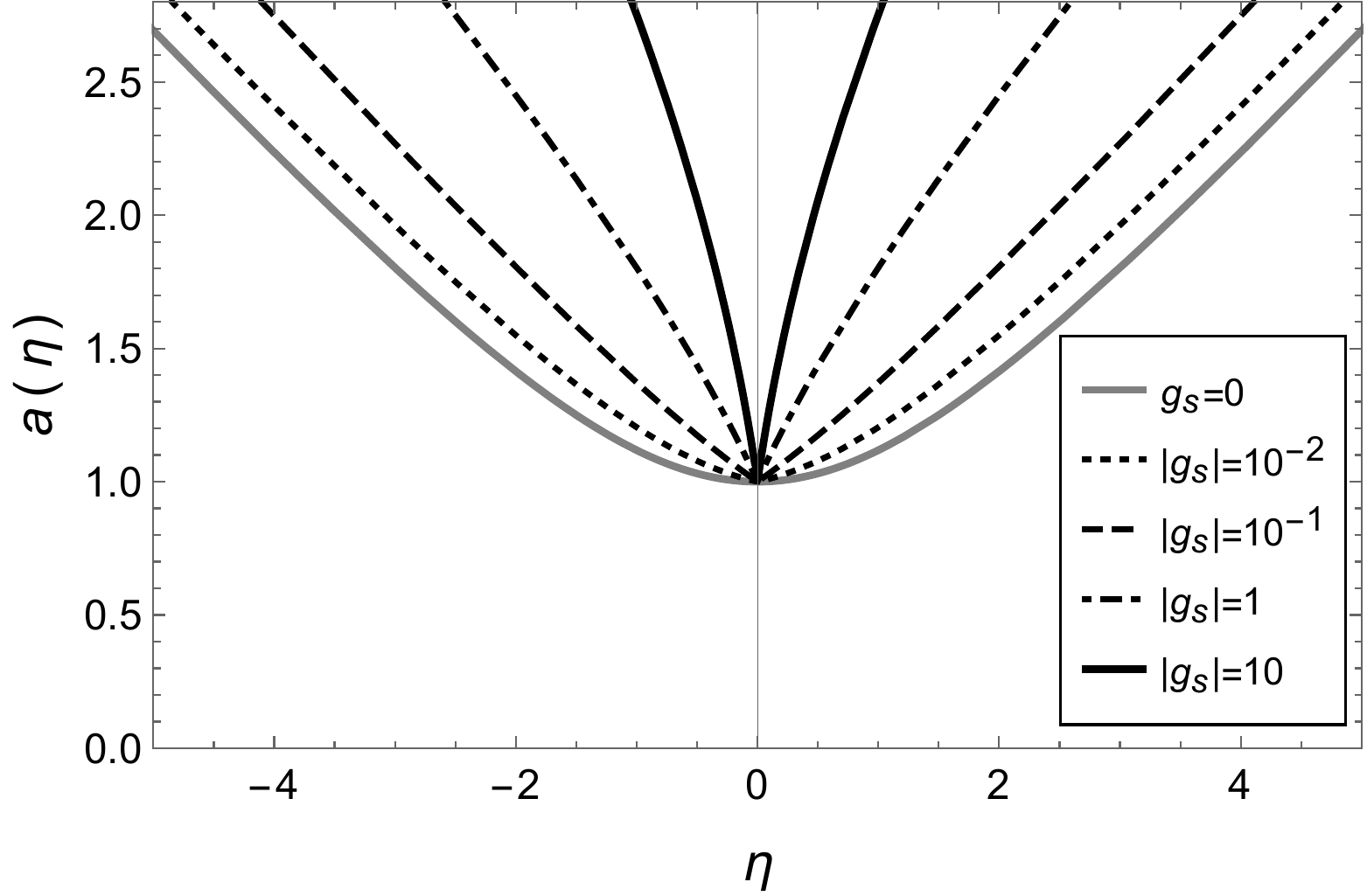}
\caption{Analytical scale factor in quantum HL theory as a function of the conformal time  in the limit $g_c\to0$. 
The plots are given for $a_B=1$, $P_\varphi=0$, $\sigma=0.5$ and some representative values of $g_s$.
The plots are valid for both spatial sections, $k=\pm1$, where $g_s>0\,(g_s<0)$ for $k=1 \,(k=-1)$.}
\label{fig2}
\end{figure}
\end{center}

\begin{center}
\begin{figure}[htb!]
\subfigure[]{\includegraphics[width=7.3cm]{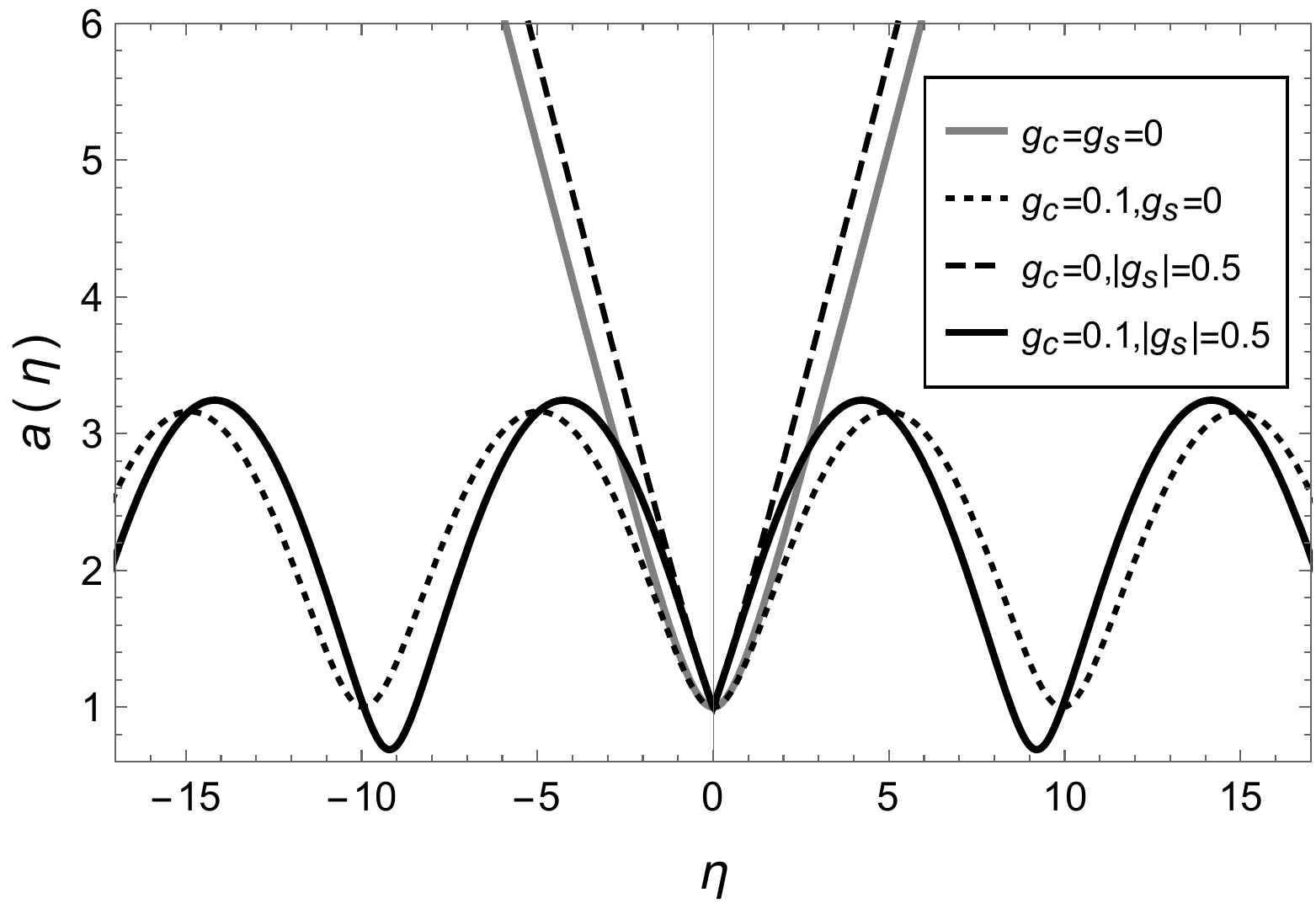}}
\subfigure[]{\includegraphics[width=7.3cm]{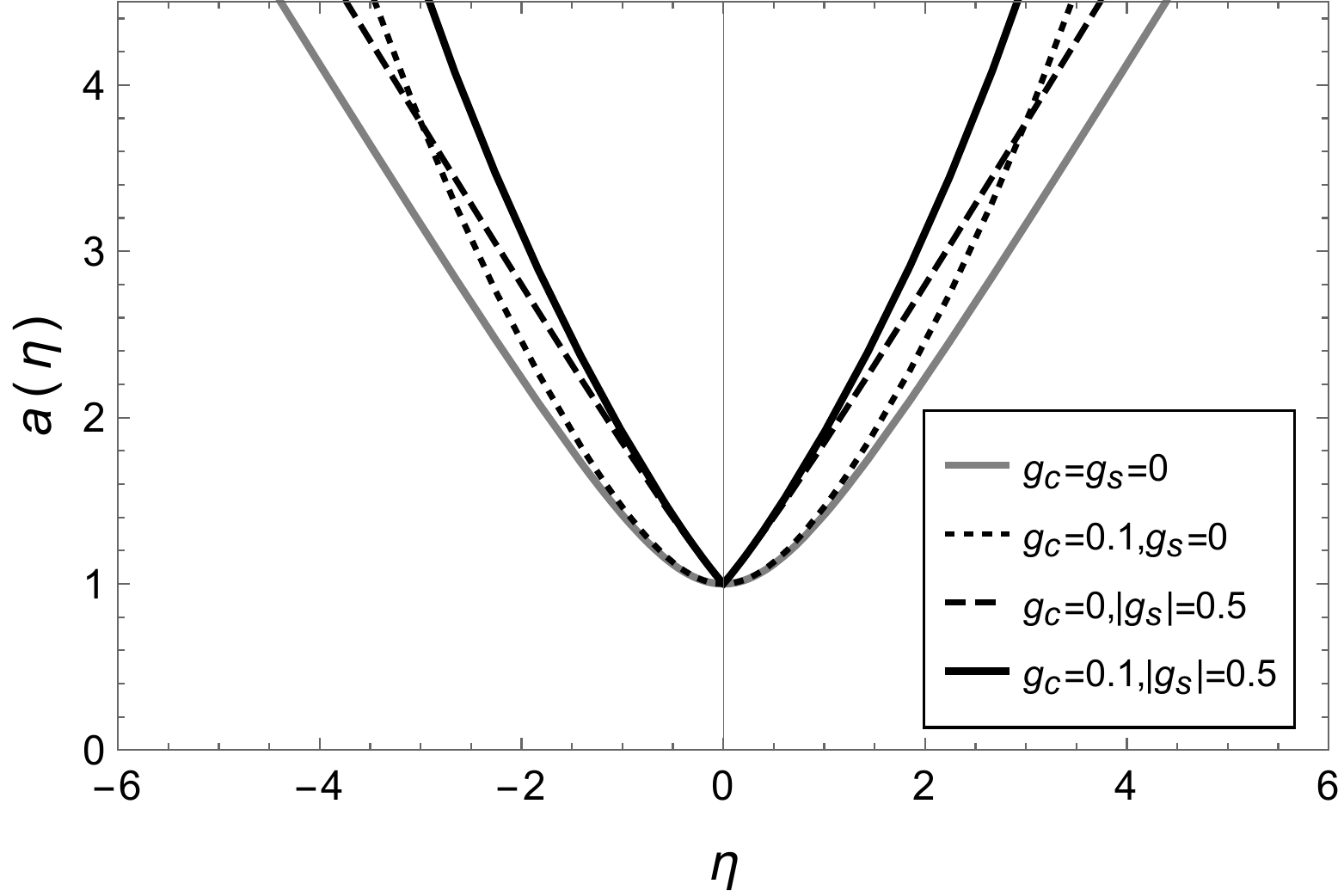}}
\caption{Analytical scale factor in quantum HL theory as a function of the conformal time for representative values of $g_c$ and $g_s$. 
The plots are given for $a_B=1$, $P_\varphi=0$, $\sigma=0.5$.
In panels (a) and (b) the spatial sections $k=1$ and $k=-1$ are considered, respectively.}
\label{fig3}
\end{figure}
\end{center}
\end{widetext}

From the solution for the wave function, Eq.~\eqref{Psigs0}, one can also obtain the quantum potential, Eq.~\eqref{Q}, which is responsible for the quantum effects.
The full analytical expression is lengthy, but one can obtain a reasonable approximation noticing from Eqs.~\eqref{aqu} that the dust fluid contribution dominates only far from the bounce. 
The quantum effects are present in the whole evolution for $k=1$ when $g_c\neq0$, but the dust fluid contribution does not significantly affect the qualitative quantum potential behavior. 
Therefore, it is a good approximation to consider Eqs.~\eqref{aqu} when $P_\varphi$ is negligible. 
From these considerations, the 
quantum potential, Eq.~\eqref{Q}, reads:
\begin{eqnarray}\label{Qsol}
\!\!\!\!\!\!\!\!\!\!\!
Q(\eta)
=
\frac{a_B^2 
\left[
\sqrt{\sigma ^2-|g_s|}
-
4 a_B^2 \left(\sigma^2-|g_s|\right)
\right]
}
{a(\eta)^2},
\end{eqnarray}
where $a(\eta)$ is given by Eqs.~\eqref{aqu} for negligible $P_\varphi$.
This result is a generalization of Eq.~(22) of Ref.~\cite{Oliveira-Neto:2017yui}.
One explicitly notices that when $a_B=0$, i.e., the universe is singular, the quantum potential vanishes.

\begin{center}
\begin{figure}[htb!]
\includegraphics[width=7.3cm]{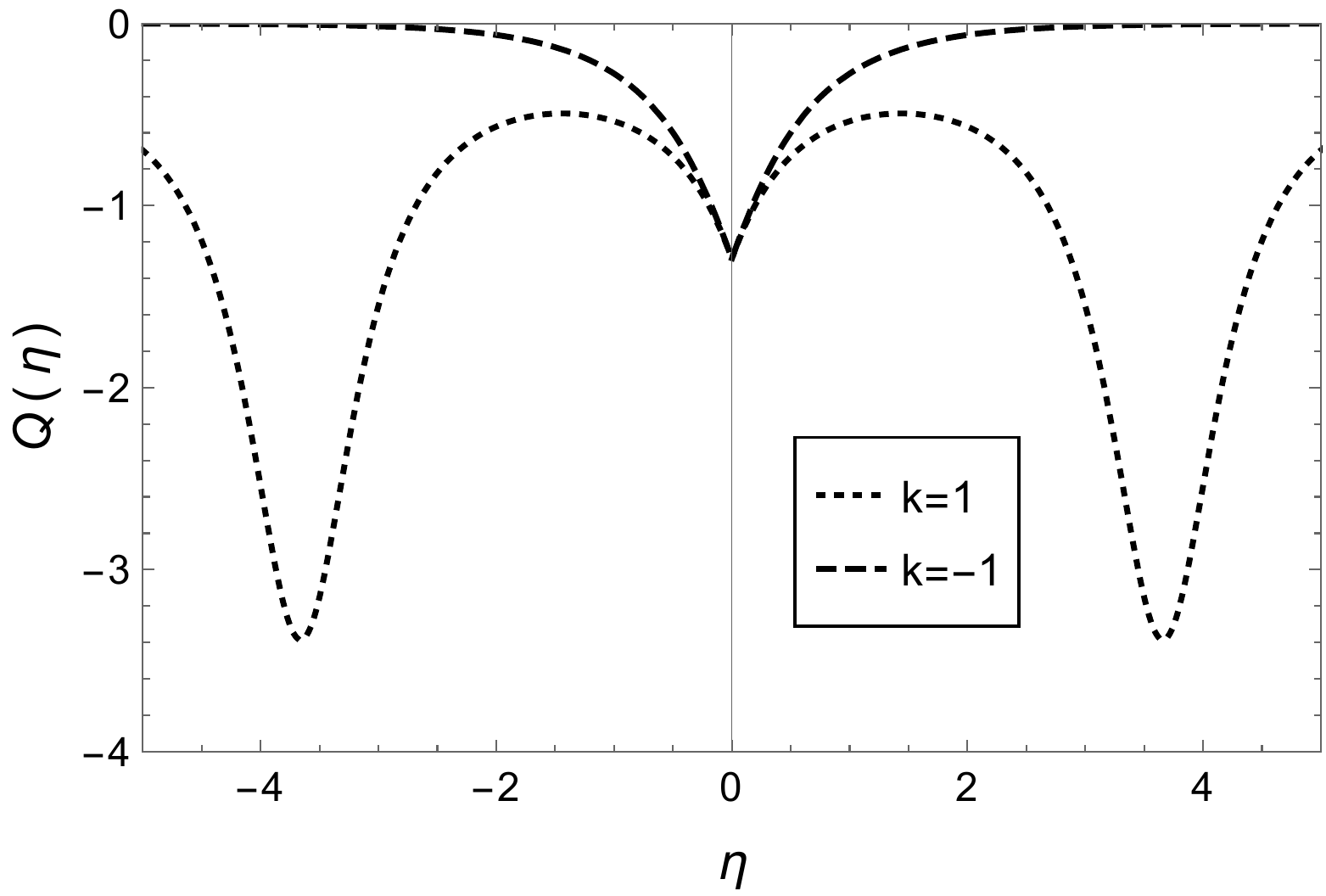}
\caption{Quantum potential in HL theory as a function of the conformal time for both $k=\pm 1$. 
The plots are given for the representative values $a_B=1$, $P_\varphi=0$ and $\sigma=g_c=|g_s|=0.5$.}
\label{fig4}
\end{figure}
\end{center}

In Fig.~\ref{fig4}, one can notice that the behavior of $Q(\eta)$ for both spatial sections is in agreement with the panels of Fig.~\ref{fig3}.
For $k=1$, the cyclic universe solutions with multiple bounces show that the quantum potential is also oscillatory and non-negligible in the whole time evolution.
On the other hand, for $k=-1$ there is a unique bounce at $\eta=0$ dominated by the $g_s$ effect, which becomes more curved as we increase the values of $g_c$ and $g_s$. 

\section{Conclusions}
\label{conclusions}

In this work I have presented a quantum minisuperspace model of FLRW cosmology in the framework HL theory of gravity.
For this purpose, I have considered the HL theory in its projectable version and without detailed balance condition.  
The Universe is filled with non-interacting radiation and dust fluids, which were introduced using the Schutz approach.
Canonical quantization is then performed and a Wheeler-DeWitt equation is obtained, Eq.~\eqref{WDWraddustalpha}.
Performing a separation of variables and a suitable choice of parameters, an analytical solution of Eq.~\eqref{WDWraddustalpha} for the wave function was first obtained in this context for nonzero values of $g_c$, $g_r$ and $g_s$ present in the HL Hamiltonian, Eq.~\eqref{Hraddust}. 
I have set $g_\Lambda=0$, which will be considered in a future work.

From the solution of the wave function, I have considered the dBB interpretation of quantum mechanics in order to derive analytical solutions for the scale factor, given by Eqs.~\eqref{aqu}.
These are quantum bounce solutions in the cases of closed and open FLRW quantum cosmologies in HL theory.
The closed universe solutions are cyclic and with multiple bounces, which avoid the big bang and big crunch singularities.
On the other hand, the open universe solutions are contracting/expanding unique bounce solutions, which get more curved as one increases the values of the $g_c$ and $g_s$ HL parameters.

I have also obtained the quantum potential, Eq.~\eqref{Qsol}, whose qualitative behavior confirms its role in the quantum evolution of each solution, i.e., that it
is the responsible for the avoidance of the singularities. 

Classical results were also presented in Sec.~\ref{classical}, which were important in order to identify $\bar{\sigma}$ and construct the quantum results with a well-defined classical limit.
In the Appendix~\ref{appendix}, an ansatz for the wave function was derived in the case where only the radiation fluid degree of freedom and the $g_s$ parameter of HL gravity are nonzero, which confirms the value of $\bar{\sigma}$.

As future perspectives, these results will be important to  study gravitational particle production and baryogenesis in this context.
On the other hand, it would also be interesting to consider a stiff matter fluid ($\omega=1$) in this context as well $g_\Lambda\neq0$, as aforementioned.


\section*{ACKNOWLEDGMENTS}

I thank Yves E. Chifarelli for useful discussions during the elaboration of this paper.

\


\appendix

\section{Ansatz solution for a quantum FLRW universe filled with radiation and nonzero $g_s$ HL parameter} 
\label{appendix}

In this section, I consider the quantum cosmology for HL theory in the particular case where $g_c=g_r=0$ and $g_s\neq0$ (apart from $g_\Lambda$, which is neglected in the whole paper) filled with a radiation fluid. 
This case corresponds to a particular case of Eq.~\eqref{WDWraddustalpha}, which reads:
\begin{eqnarray}\label{WDWradgsalpha}
\!\!\!\!\!\!\!\!\!
i\partial_\eta \Psi(a,\eta)
&=&
\left(
-\frac{1}{4}\partial_a^2 
+ \frac{\alpha}{4a}\partial_a  
- \frac{g_s k}{a^2}
\right)\Psi(a,\eta).
\end{eqnarray}
A solution for this equation can be obtained from the following ansatz:
\begin{eqnarray}
\Psi_{\rm ansatz}(a,\eta)
=
a^{\alpha/2}f(\eta)\,e^{-g(\eta)a^2},
\end{eqnarray}
where $f(\eta)$ and $g(\eta)$ are arbitrary functions of $\eta$.
From Sec.~\ref{quantumcosmology}, the quantum analytical solution for this case, where the particular choice $\alpha=-1+\sqrt{1+16 g_s k}$ was considered, is given by:
\begin{eqnarray}\label{aqugc0Pvarphi0}
\!\!\!\!\!\!\!\!\!\!\!
a(\eta)=a_B\sqrt{1+ 
2 \sqrt{g_s k}\, |\eta|+ \sigma^2\eta^2 }.
\end{eqnarray}
The expressions for $f(\eta)$ and $g(\eta)$ of the wave-function ansatz which reproduce this analytical solution read:
\begin{eqnarray}
f(\eta)=
\frac{1}{\sqrt{1+i \bar{\sigma}\eta}}
\qquad,\qquad
g(\eta)=
\frac{\bar{\sigma}}{1+i\bar{\sigma} \eta}.
\end{eqnarray}
where $\bar{\sigma}=\sqrt{\sigma^2-g_s k}+i\sqrt{g_s k}$.
In terms of these functions, the  ansatz results:
\begin{eqnarray}
\Psi_{\rm ansatz}(a,\eta)
\propto
a^{(-1+\sqrt{1+16 g_s k})/2}
\frac{e^{-\frac{\bar{\sigma}a^2}{1+i\bar{\sigma} \eta} }}{\sqrt{1+i \bar{\sigma}\eta}},
\end{eqnarray}
From the ansatz, the initial condition at $\eta=0$ reads:
\begin{eqnarray}
\Psi_{\rm ansatz}(a,0)
\propto
a^{(-1+\sqrt{1+16 g_s k})/2}
e^{-\bar{\sigma}a^2 }.
\end{eqnarray}
Comparing this result with Eqs.~\eqref{psidustgr} and~\eqref{psiref}, this is exactly the initial dependence on the scale factor $a$.
One can also notice the presence of the parameter $\bar{\sigma}$ in the exponential. 
Therefore, Eq.~\eqref{psi0} is the appropriate initial condition, noticing that the term $a^{(-1+\sqrt{1+16 g_s k})/2}$ was factorized by a previous change of variables. 


\end{document}